\documentclass[a4paper,12pt]{article}
\usepackage{graphicx}
\usepackage{bm}
\usepackage{amstext,amsmath,amssymb}
\usepackage{mathptmx}
\usepackage[top=3cm,bottom=3cm,left=1.5cm,right=1.5cm,asymmetric]{geometry}
\usepackage{rotating}
\usepackage{hyperref}
\usepackage{cite}
\usepackage{array}
\usepackage{tabularx}
\DeclareMathOperator{\sech}{sech}

\begin{document}

\title{A Transformation Method to Construct Family of Exactly Solvable Potentials in Quantum Mechanics}
\author{\textbf{Nabaratna Bhagawati}$^{1}$, \textbf{N Saikia}$^{2}$ and \textbf{N Nimai Singh}$^{3}$
\\$^{1,3}$Department of Physics, Gauhati University, Guwahati-781014, India
\\$^{2}$Department of Physics, Chaiduar College, Gohpur-784168, India
\\$^{1}$E-mail: nabaratna2008@gmail.com} 
\date{}
\maketitle

\abstract{A transformation method is applied to the second order ordinary differential equation satisfied by orthogonal polynomials to construct a family of exactly solvable quantum systems in any arbitrary dimensional space. Using the properties of orthogonal polynomials, the method transforms polynomial differential equation to D-dimensional radial Schrodinger equation which facilitates construction of exactly solvable quantum systems. The method is also applied using associated Laguerre and Hypergeometric polynomials. The quantum systems generated from other polynomials are also briefly highlighted.
}

\section{Introduction}
\label{sec1}
Schrodinger equation plays a pivotal role in modern physics as its solution gives complete information of any given non-relativistic quantum system. Along the years, many authors have tried to obtain the exact solution of Schrodinger equation for potentials of physical interest~\cite{Bhattacharjie, Fleases1, Fleases2, Khare, Dutt1, Levai, De, de, Dutt2, Ahmed1, Dong, Singh}. This is because, despite the intrinsic interest of the exactly solvable systems, these solutions can be used to get better approximated solutions for potentials which are physically interesting. To enhance the set of exactly solvable potentials, we follow a simple and compact transformation method~\cite{Ahmed1, Saikia1, Saikia2, Ahmed2, Saikia3} which comprises of a co-ordinate transformation supplemented by a functional transformation. By applying this method, we transform the second order ordinary differential equation satisfied by special functions to standard Schrodinger equation in arbitrary $D$-dimensional Euclidean space and thus try to construct as many exactly solvable potentials as possible. The method is efficient in generating both power and non-power law type spherically symmetric potentials.\\
\hspace*{0.3in}The article is organized as follows. In section~\ref{sec2}, the detailed formalism of the theory is given. In section~\ref{sec3}, the application of the method using associated Laguerre polynomial is discussed. Also the solvable potentials obtained from Hypergeometric, associated Legendre and Jacobi polynomials are tabulated. The conclusions are discussed in section~\ref{sec4}.

\section{Formalism}
\label{sec2}
We consider a second order differential equation satisfied by a special function $Q(r)$
\begin{eqnarray}
\label{eq1}
Q^{\prime\prime}(r)+M(r)Q^{\prime}(r)+J(r)Q(r)=0
\end{eqnarray}
where a prime denotes differentiation with respect to its argument. $Q(r)$ will later be identified as one of the orthogonal polynomials.
\\ The transformation method comprises of the following two steps
\begin{equation}
\label{eq2}
r\rightarrow g(r)
\end{equation}
\begin{equation}
\label{eq3}
\psi(r)=f^{-1}(r)Q(g(r))
\end{equation}
We implement the above prescription to equation~(\ref{eq1}) and obtain
\begin{eqnarray}
\label{eq4}
&&\psi^{\prime\prime}(r)+\left(\frac{d}{dr}\ln\frac{f^{2}(r)\exp(\int M(g)dg)}{g^{\prime}(r)}\right)\psi^{\prime}(r)+\nonumber \\
&&\left(\frac{f^{\prime\prime}(r)}{f(r)}-\frac{g^{\prime\prime}(r)}{g^{\prime}(r)}\frac{f^{\prime}(r)}{f(r)}+g^{\prime}(r)M(g)\frac{f^{\prime}(r)}{f(r)}+g^{\prime 2}J(g)\right)\psi(r)=0
\end{eqnarray}
The radial Schrodinger equation in $D$-dimensional Euclidean space is($\hbar=1=2m$)
\begin{eqnarray}
\label{eq5}
\psi^{\prime\prime}(r)+\frac{(D-1)}{r}\psi^{\prime}(r)+\left(E_{n}-V(r)-\frac{\ell(\ell+D-2)}{r^2}\right)\psi(r)=0
\end{eqnarray}
Consistency of equations~(\ref{eq4}) and~(\ref{eq5}) demand that 
\begin{equation}
\label{eq6}
\frac{d}{dr}\ln\frac{f^{2}(r)\exp(\int M(g)dg)}{g^{\prime}(r)}=\frac{(D-1)}{r}
\end{equation}
which fixes the form of $f(r)$ as
\begin{equation}
\label{eq7}
f(r)=N r^{\frac{(D-1)}{2}}g^{\prime\frac{1}{2}}\left(\exp\left(-\int M(g)dg\right)\right)^{\frac{1}{2}}
\end{equation}
where $N$ is the integration constant and plays the role of the normalization constant of the wavefunctions.
\\Using (\ref{eq6}) and (\ref{eq7}) in equation (\ref{eq4}) yields
\begin{align}
\label{eq8}
\frac{\psi^{\prime\prime}(r)}{\psi(r)}+\frac{(D-1)}{r}\frac{\psi^{\prime}(r)}{\psi(r)}=-\frac{1}{2}\{g,r\}+\frac{g^{\prime2}(r)}{4}\left[M^{2}(g)+2M^{\prime}(g)-4J(g)\right]-\frac{(D-1)(D-3)}{4r^2}
\end{align}
where the Schwartzian derivative symbol~\cite{Hille}, $\{g,r\}$ is defined as
\begin{equation*}
\{g,r\}=\frac{g^{\prime\prime\prime}(r)}{g^{\prime}(r)}-\frac{3}{2}\frac{g^{\prime\prime2}(r)}{g^{\prime2}(r)}
\end{equation*}
From equations (\ref{eq3}) and (\ref{eq7}), the expression for nomalizable wavefunction is
\begin{equation}
\label{eq9}
\psi(r)=N r^{-\frac{(D-1)}{2}}g^{\prime-\frac{1}{2}}\left(\exp\left(\int M(g)dg\right)\right)^{\frac{1}{2}}Q(g(r))
\end{equation}
The radial wavefunction $\psi(r)=\frac{u(r)}{r}$ has to satisfy the boundary condition $u(r)=0$, in order to rule out singular solutions~\cite{Khelashvili}.\\
Expression (\ref{eq8}) can be cast in the standard Schrodinger equation form (equation (\ref{eq5})) if we can write
\begin{equation}
\label{eq10}
-(E_n-V(r))=-\frac{1}{2}\{g,r\}+\frac{g^{\prime2}(r)}{4}\left[M^{2}(g)+2M^{\prime}(g)-4J(g)\right]-\frac{(D-1)(D-3)}{4r^2}
\end{equation}
Once we choose a particular orthogonal polynomial, $Q(g)$, to construct an exact solution of the Schrodinger equation, the characteristic functions of the polynomial $M(g)$, $J(g)$ get specified. We have to choose one or more than one terms containing the function $g(r)$ in expression (\ref{eq10}) and put it equal to a constant to get the energy eigenvalues $E_n$. The procedure is worked out in detail for Laguerre and Hypergeometric polynomials in the next section.\\
It is interesting to note that when the generated potential is purely non-power law, the potential given by expression (\ref{eq10}) has a term $\frac{(D-1)(D-3)}{4r^2}$ which behaves as constant background attractive inverse square potential in any arbitrary dimension except for dimensions 1 and 3. For power law cases, this background potential and the potential coming from Schwartzian derivative unite to give the correct centrifugal barrier potential in arbitrary dimensions.

\section{Application of the transformation method}
\label{sec3}
\subsection{Construction of exactly solvable potentials from associated Laguerre polynomial}
Identifying
\begin{align}
\label{eq11}
Q(g(r))=L_n^\alpha(g)
\end{align}
as the associated Laguerre polynomial and its characteristic functions $M(g)$ and $J(g)$ are
\begin{align}
\label{eq12}
M(g)=\frac{\alpha+1-g}{g}
\end{align}
\begin{align}
\label{eq13}
J(g)=\frac{n}{g}
\end{align}
Using equations (\ref{eq11}),(\ref{eq12}) and~(\ref{eq13}) in equation (\ref{eq8}) yields
\begin{align}
\label{eq14}
\frac{\psi^{\prime\prime}(r)}{\psi(r)}+\frac{(D-1)}{r}\frac{\psi^{\prime}(r)}{\psi(r)}=\frac{1}{4}(\alpha^2-1)\frac{g^{\prime2}}{g^2}-\frac{1}{2}(2n+\alpha+1)\frac{g^{\prime2}}{g}+\frac{g^{\prime2}}{4}-\frac{1}{2}\frac{g^{\prime\prime\prime}}{g^{\prime}}+\frac{3}{4}\frac{g^{\prime\prime2}}{g^{\prime2}}-\frac{(D-1)(D-3)}{4r^2}
\end{align}
and equation (\ref{eq9}) yields
\begin{align}
\label{eq15}
\psi(r)=Nr^{-\frac{(D-1)}{2}}g^{\prime-\frac{1}{2}}g^{\frac{\alpha+1}{2}}\exp(-\frac{g}{2})L_n^\alpha(g)
\end{align}
To convert equation (\ref{eq14}) into a standard stationary state Schrodinger equation, we make one or more terms of the right hand side of equation (\ref{eq14}) a constant quantity. This enables us to get the energy eigenvalues $E_n$, the functional form of $g(r)$ and subsequently potential $V(r)$ and wavefunction $\psi(r)$.\\ \\
\hspace*{0.3in}(\romannumeral1) As a first case, let us choose
\begin{align}
\label{eq16}
\frac{g^{\prime2}}{g^2}=c_1^2
\end{align}
where $c_1^2$ is a real positive constant independent of $r$. Equation (\ref{eq16}) gives the functional form of $g(r)$ as
\begin{align}
\label{eq17}
g(r)=A_1\exp(-c_1r)
\end{align}
where $A_1$ is an integration constant and for normalizability condition we consider here only the negative sign in the exponential. Using the value of $g(r)$ in equation (\ref{eq14}) yields
\begin{align}
\label{eq18}
E_n=-\frac{c_1^2\alpha^2}{4}
\end{align}
\begin{align}
\label{eq19}
V(r)=A_1c_1^2\exp(-c_1r)\left(\frac{A_1}{4}\exp(-c_1r)-\frac{2n+\alpha+1}{2}\right)-\frac{(D-1)(D-3)}{4r^2}
\end{align}
and from equation (\ref{eq15}) we obtain
\begin{align}
\label{eq20}
\psi(r)=Nr^{-\frac{(D-1)}{2}}\exp(-\frac{c_1\alpha r}{2})\exp(-\frac{A_1\exp(-c_1r)}{2})L_n^{\alpha-1}(\exp(-c_1r))
\end{align}
To express energy eigenvalues in terms of the quantum number $n$, we choose
\begin{align}\nonumber
\frac{2n+\alpha+1}{2}=\beta
\end{align}
a constant independent of $n$, which gives $\alpha=2\beta-2n-1$ and $\beta\geq n+\frac{1}{2}$. This yields energy eigenvalues, potential and energy eigenfunction as $(A_1=1)$
\begin{align}
\label{eq21}
E_n=-\frac{c_1^2}{4}(2\beta-2n-1)^2
\end{align}
\begin{align}
\label{eq22}
V(r)=c_1^2\exp(-c_1r)\left(\frac{1}{4}\exp(-c_1r)-\beta\right)-\frac{(D-1)(D-3)}{4r^2}
\end{align}
and
\begin{align}
\label{eq23}
\psi(r)=Nr^{-\frac{(D-1)}{2}}\exp\left(-(2\beta-2n-1)\frac{c_1r}{2}\right)\exp\left(-\frac{\exp(-c_1r)}{2}\right)L_n^{2\beta-2n-2}(\exp(-c_1r))
\end{align}
The potential given by expression (\ref{eq22}) is non-power law and as our formalism suggests, it has an inverse square potential term in spaces where the dimensionality is other than 1 and 3.\\ \\
\hspace*{0.3in}(\romannumeral2) Continuing the procedure to construct exactly solvable quantum system we consider second term $\frac{g^{\prime2}}{g}$ of expression (\ref{eq14}) to be constant independent of $r$.\\
i.e.,
\begin{align}
\label{eq24}
\frac{g^{\prime2}}{g}=c_2^2
\end{align}
we get the functional form of $g(r)$ as
\begin{align}
\label{eq25}
g(r)=\frac{c_2^2}{4}r^2
\end{align}
Equations (\ref{eq14}) and (\ref{eq25}) yield
\begin{align}
\label{eq26}
E_n=\frac{1}{2}(2n_r+\alpha+1)c_2^2
\end{align}
\begin{align}
\label{eq27}
V(r)=\frac{c_2^4}{16}r^2+\left(\alpha^2-\frac{1}{4}-\frac{(D-1)(D-3)}{4}\right)\frac{1}{r^2}
\end{align}
and
\begin{align}
\label{eq28}
\psi(r)=Nr^{\alpha+1-\frac{D}{2}}\exp(-\frac{c_2^2}{8}r^2)L_{n_r}^\alpha(\frac{c_2^2}{4}r^2)
\end{align}
To get the correct centrifugal barrier term in $D$-dimensional Euclidean space, we have to identify the co-efficent of $\frac{1}{r^2}$ in expression (\ref{eq27}) to be $\ell(\ell+D-2)$, which fixes the value of $\alpha$ as
\begin{align}
\label{eq29}
\alpha=\ell+\frac{D-2}{2}
\end{align}
Identifying
\begin{align}
\label{eq30}
\frac{c_2^2}{2}=\omega
\end{align}
Expressions (\ref{eq26}), (\ref{eq29}) and (\ref{eq30}) yield energy eigenvalues as
\begin{align}
\label{eq31}
E_n=\omega(n+\frac{D}{2})
\end{align}
where the principal quantum number, $n$ is $(2n_r+\ell)$.\\
\hspace*{0.3in} From equations (\ref{eq26}) and (\ref{eq27}) we get the potential and eigenfunction as
\begin{align}
\label{eq32}
V(r)=\frac{1}{4}\omega^2r^2+\frac{\ell(\ell+D-2)}{r^2}
\end{align}
and
\begin{align}
\label{eq33}
\psi(r)=Nr^\ell\exp(-\frac{\omega r^2}{4})L_{\frac{1}{2}(n-\ell)}^{\ell+\frac{D-2}{2}}(\frac{\omega r^2}{2})
\end{align}
respectively. \\ \\
\hspace*{0.3in}(\romannumeral3) Proceeding in a similar way, let
\begin{align}
\label{eq34}
g^{\prime2}(r)=c_3^2
\end{align}
This gives
\begin{align}
\label{eq35}
g(r)=c_3r
\end{align}
As required for the normalizability of the wavefunction we have taken the positive sign. Equations (\ref{eq14}) and (\ref{eq35}) yield
\begin{align}
\label{eq36}
E_n=-\frac{c_3^2}{4}
\end{align}
\begin{align}
\label{eq37}
V(r)=-\frac{c_3}{2}(2n_r+\alpha+1)\frac{1}{r}+\frac{1}{4r^2}\left(\alpha^2-1-(D-1)(D-3)\right)
\end{align}
and
\begin{align}
\label{eq38}
\psi(r)=Nr^{\frac{1}{2}(\alpha-D+2)}\exp(-\frac{c_3}{2}r)L_n^{\alpha+1}(c_3r)
\end{align}
To get the correct form of centrifugal barrier term in $D$-dimensional Euclidean space, we have to identify the coefficient of $\frac{1}{r^2}$ in expression (\ref{eq37}) to be $\ell(\ell+D-2)$, which fixes the value of $\alpha$ as
\begin{align}
\label{eq39}
\alpha+1=2\ell+D-1
\end{align}
Further, to get the energy eigenvalues in terms of the quantum number $n$, we identify the co-efficient of $\frac{1}{r}$ in equation (\ref{eq37}) to be $e^2$, which in atomic unit is 2, i.e.,
\begin{align}
\label{eq40}
c_3=\frac{4}{2n_r+\alpha+1}
\end{align}
Equations (\ref{eq36}), (\ref{eq39}) and (\ref{eq40}) yield the energy eigenvelues
\begin{align}
\label{eq41}
E_n=-\frac{1}{n^2}
\end{align}
where the principal quantum number, $n$ for the $D$-dimensional case is $n=n_r+\ell+\frac{D-1}{2}$, and reduces to the usual, $n=n_r+\ell+1$ when $D=3$. Now, the potential becomes
\begin{align}
\label{eq42}
V(r)=-\frac{2}{r}+\frac{\ell(\ell+D-2)}{r^2}
\end{align}
and the wavefunction
\begin{align}
\label{eq43}
\psi(r)=Nr^\ell \exp(-\frac{r}{n})L_n^{2\ell+D-1}(\frac{2r}{n})
\end{align}\\
\hspace*{0.3in}(\romannumeral4) As a fourth choice, let
\begin{align}
\label{eq44}
\frac{g^{\prime\prime2}}{g^{\prime2}}=c_4^2
\end{align}
We get the functional form of $g(r)$ similar to that of the first choice which gives us a similar type of quantum system as obtained in that case.\\ \\

\begin{sidewaystable}[p]
 \centering
 \def\~{\hphantom{0}}
\hsize\textheight
 \caption{Summary of the Quantum Systems obtained from Associated Laguerre Polynomial $L_n^\alpha(g)$}
\begin{tabular*}{\textwidth}{@{}l@{\extracolsep{\fill}}
l@{\extracolsep{\fill}}l@{\extracolsep{\fill}}l@{\extracolsep{\fill}}}
\hline
 ${g(r)}$ & $E_n$ & V(r) & $\psi(r)$ \\ 
\hline

$\exp(-c_1r)$ & $-\frac{c_1^2}{4}(2\beta-2n-1)^2$ & $c_1^2\exp(-c_1r)\left(\frac{1}{4}\exp(-c_1r)-\beta\right)$ & $Nr^{-\frac{(D-1)}{2}}\exp\left(-(2\beta-2n-1)\frac{c_1r}{2}\right)$\\
$$ & $$ & $-\frac{(D-1)(D-3)}{4r^2}$ & $\exp\left(-\frac{\exp(-c_1r)}{2}\right)L_n^{2\beta-2n-2}(\exp(-c_1r))$\\ \\
$$ & $$ & $where, \frac{2n+\alpha+1}{2}=\beta$ & $$\\ \\

$\frac{c_2^2}{4}r^2$ & $\omega(n+\frac{D}{2})$ & $\frac{1}{4}\omega^2r^2+\frac{\ell(\ell+D-2)}{r^2}$ & $Nr^\ell\exp(-\frac{\omega r^2}{4})L_{\frac{1}{2}(n-\ell)}^{\ell+\frac{D-2}{2}}(\frac{\omega r^2}{2})$\\ \\
$$ & $where,\frac{c_2^2}{2}=\omega$ & $$ & $where,\alpha=\ell+\frac{D-2}{2}$\\ \\

$c_3r$ & $-\frac{1}{n^2}$ & $-\frac{2}{r}+\frac{\ell(\ell+D-2)}{r^2}$ & $Nr^\ell \exp(-\frac{r}{n})L_n^{2\ell+D-1}(\frac{2r}{n})$\\ \\
$$ & $where,n=n_r+\ell+\frac{D-1}{2}$ & $where,c_3=\frac{4}{2n_r+\alpha+1}$ & $where,\alpha+1=2\ell+D-1$\\ 

\hline 
\end{tabular*}
\label{Tab1}
\end{sidewaystable}
The summary of the constructed exactly solvable systems are listed in Table~\ref{Tab1}

\subsection{Construction of exactly solvable potentials from Hypergeometric function}
Identifying
\begin{align}
\label{eq50}
Q(g(r))={_2}F_1(\alpha,\beta,\gamma;g)
\end{align}
as the hypergeometric function and its characteristic functions $M(g)$ and $J(g)$ are
\begin{align}
\label{eq51}
M(g)=\frac{\gamma-(\alpha+\beta+1)g}{g(1-g)}
\end{align}
\begin{align}
\label{eq52}
J(g)=-\frac{\alpha\beta}{g(1-g)}
\end{align}
Using equations (\ref{eq50}),(\ref{eq51}) and (\ref{eq52}) in equation (\ref{eq8}) yield
\begin{align}\nonumber
\label{eq53}
&\frac{\psi^{\prime\prime}(r)}{\psi(r)}+\frac{(D-1)}{r}\frac{\psi^{\prime}(r)}{\psi(r)}=\left(\frac{\gamma(\gamma-\alpha-\beta-1)}{2}+\alpha\beta\right)\frac{g^{\prime2}}{g}+\frac{\gamma(\gamma-2)}{4}\frac{g^{\prime2}}{g^2}&\\
&+\left(\frac{\gamma(\gamma-\alpha-\beta-1)}{2}+\alpha\beta\right)\frac{g^{\prime2}}{(1-g)}+\left(\frac{(\alpha+\beta-\gamma)^2-1}{4}\right)\frac{g^{\prime2}}{(1-g)^2}-\frac{1}{2}\frac{g^{\prime\prime\prime}}{g^{\prime}}+\frac{3}{4}\frac{g^{\prime\prime2}}{g^{\prime2}}-\frac{(D-1)(D-3)}{4r^2}&
\end{align}
and equation (\ref{eq9}) yield
\begin{align}
\label{eq54}
\psi(r)=Nr^{-\frac{(D-1)}{2}}g^{\prime-\frac{1}{2}}g^{\frac{\gamma}{2}}(1-g)^{\frac{\alpha+\beta-\gamma+1}{2}}{_2}F_1(\alpha,\beta,\gamma;g)
\end{align}
To put equation (\ref{eq53}) into standard stationary state Schrodinger equation and to generate exactly solvable quantum systems, we follow the same procedure of equating different terms of the right hand side of equation (\ref{eq53}) to a constant. We summarize the different quantum systems thus obtained in Table~\ref{Tab2}.

\begin{sidewaystable}[p]
 \centering
 \def\~{\hphantom{0}}
\hsize\textheight
 \caption{Summary of the Quantum Systems obtained from Hypergeometric Function ${_2}F_1(\alpha,\beta,\gamma;g)$}
\begin{tabular*}{\textwidth}{@{}l@{\extracolsep{\fill}}l@{\extracolsep{\fill}}
l@{\extracolsep{\fill}}l@{\extracolsep{\fill}}l@{\extracolsep{\fill}}}
\hline
${Relation}$ & ${g(r)}$ & $E_n$ & V(r) & $\psi(r)$ \\ 
\hline

$\frac{g^{\prime2}}{g}$ & $\frac{p_1^2}{4}r^2$ & $-2(-2n(\beta+1)+(\ell+\frac{D}{2})$ & $\frac{r^2}{(1-r^2)}(-4n(\beta+1)+$ & $r^{\ell} (1-r^2)^{\frac{-n+\beta-\ell-\frac{D}{2}+2}{2}}$\\

$=p_1^2$ & $$ & $(\ell+\frac{D}{2}+n-\beta-2))$ & $(2\ell+D)(2\ell+D+2n-2\beta-4)$ & ${_2}F_1(-n,\beta+1,\ell+\frac{D}{2};r^2)$\\
$$ & $$ & $$ & $+\frac{(-n+\beta-\ell-\frac{D}{2}+1)^2-1}{4(1-r^2)})+\frac{\ell(\ell+D-2)}{r^2}$ & $$\\ \\ 
  $$ & $$ & $$ & $where,c=\ell+\frac{D}{2}; \hspace{0.05in} taking, p_1=2$ & $$ \\ \\

$\frac{g^{\prime2}}{g^2}$ & $A_2\exp(-p_2r)$ & $-p_2^2\left(\frac{\beta_1^2-(n+1)^2}{2(n+1)}\right)^2$ & $-\frac{\beta_1^2A_2p_2^2\exp(-p_2r)}{1-A_2\exp(-p_2r)}-\frac{(D-1)(D-3)}{4r^2}$ & $r^{-\frac{D-1}{2}}(A_2\exp(-p_2r))^{\frac{\beta_1^2-(n+1)^2}{2(n+1)}}(1-A_2\exp(-p_2r))$ \\
$=p_2^2$ & $$ & $$ & $$ & ${_2}F_1(-n,\frac{\beta_1^2+n+1}{n+1},\frac{\beta_1^2-n^2-n}{n+1};A_2\exp(-p_2r))$ \\ \\
  $$ & $$ & $where,$ & $with\hspace{0.05in} condition,$ & $$\\
  $$ & $$ & $(n+1)(n-1+\gamma)=\beta_1$ & $\alpha+\beta-\gamma+1=\pm1$ & $$\\ \\

$\frac{g^{\prime2}}{(1-g)^2}$ & $1-$ & $-p_3^2(\frac{\gamma_1^2+(n+1)^2}{2(n+1)})^2$ & $\frac{p_3^2A_3\gamma_1^2\exp(-p_3r)}{(1-A_3\exp(-p_3r))}-\frac{(D-1)(D-3)}{4r^2}$ & $r^{-\frac{(D-1)}{2}}(1-A_3\exp(-p_3r))\exp(-p_3\frac{\gamma_1^2+(n+1)^2}{2(n+1)}r)$ \\[-1ex]
$=p_3^2$ & $A_3\exp(-p_3r)$ & $$ & $$ & $_2F_1(-n,\frac{n-(\gamma_1^2-1)}{n+1},2;(1-\exp(-p_3r)))$\\ \\
  $$ & $$ & $where,$ & $with, \gamma=1$\\
  $$ & $$ & $-\beta(n+1)+n+1=\gamma_1^2$ & $$ & $$ \\ \\
  
$\frac{g^{\prime\prime2}}{g^{\prime2}}$ & $A_4\exp(-p_4r)$ & $-p_4^2\left(\frac{\delta^2-(n+1)^2}{2(n+1)}\right)$ & $-\frac{\delta^2A_4p_4^2\exp(-p_4r)}{1-A_4\exp(-p_4r)}-\frac{(D-1)(D-3)}{4r^2}$ & $r^{-\frac{D-1}{2}}(A_4\exp(-p_4r))^{\frac{\delta^2-(n+1)^2}{2(n+1)}}(1-A_4\exp(-p_4r))$\\
$=p_4^2$ & $$ & $$ & $$ & ${_2}F_1(-n,\frac{\delta^2+n+1}{n+1},\frac{\delta^2-n^2-n}{n+1};A_4\exp(-p_4r))$\\ \\
 $$ & $$ & $where,$ & $with\hspace{0.05in} condition,$ & $$\\
 $$ & $$ & $(n+1)(n-1+\gamma)=\delta$ & $ \alpha+\beta-\gamma+1=\pm1$ & $$\\
\hline 
\footnote{In each case, we have taken $\alpha=-n$ to make ${_2}F_1(\alpha,\beta,\gamma;g)$ a polynomial.}
\end{tabular*}
\label{Tab2}
\end{sidewaystable}

\hspace*{0.3in} In a similar way, by identifying $Q(g)$ as another orthogonal polynomial and applying the above mentioned procedure, different exactly solvable potentials can be obtained. We have listed a few of them in Table~\ref{Tab3}.

\begin{sidewaystable}[p]
 \centering
 \def\~{\hphantom{0}}
\hsize\textheight
 \caption{Quantum Systems obtained from other Orthogonal Polynomials}
\begin{tabular*}{\textwidth}{@{}l@{\extracolsep{\fill}}l@{\extracolsep{\fill}}
l@{\extracolsep{\fill}}l@{\extracolsep{\fill}}l@{\extracolsep{\fill}}}
\hline
${Relation}$ & ${g(r)}$ & $E_n$ & V(r) & $\psi(r)$ \\ 
\hline

$\frac{g^{\prime2}}{1-g^2}$ & $\sin{p_1r}$ & $-p_1^2(m^2-n(n+1)-\frac{1}{2})$ & $C_m^2\tan^2{p_1r}-\frac{(D-1)(D-3)}{4r^2}$ & $Nr^{-\frac{D-1}{2}}\cos^{\frac{1}{2}}{p_1r}P_n^m(\sin{p_1r})$\\
$=p_1^2$ & $$ & $$ & $$ & $$\\ \\ 
  $$ & $$ & $$ & $where,p_1^2(m^2-\frac{1}{4})=C_m^2$ & $$ \\ \\

$\frac{g^{\prime2}}{(1-g^2)^2}$ & $\tanh{p_2r}$ & $-(m-1)^2p_2^2$ & $-n(n-1)p_2^2\sech^2{p_2r}-\frac{(D-1)(D-3)}{4r^2}$ & $Nr^{-\frac{D-1}{2}}P_{n-1}^{m-1}(\tanh{p_2r})$\\
$=p_2^2$ & $$ & $$ & $$ & $$ \\ \\
 
$\frac{g^{\prime2}}{1-g^2}$ & $\sin{p_3r}$ & $\frac{1}{4}((\alpha-\beta)^2-4n(n+\alpha+\beta+1)$ & $c_1^2\tan^2{p_3r}+c_2^2\sec{p_3r}\tan{p_3r}$ & $Nr^{-\frac{D-1}{2}}(\cos{p_3r})^{\frac{\alpha+\beta+1}{2}}$\\
$=p_3^2$ & $$ & $-2(\alpha+\beta)-2)p_3^2$ & $-\frac{(D-1)(D-3)}{4r^2}$ & $(\frac{1+\sin{p_3r}}{1-\sin{p_3r}})^{\frac{\beta-\alpha}{4}}P_n^{(\alpha,\beta)}(\sin{p_3r})$ \\ \\
  $$ & $$ & $$ & $with, \frac{(\alpha^2-\beta^2)}{2}p_3^2=c_1^2;$ \\
  $$ & $$ & $$ & $ \hspace{0.5in}\frac{(\alpha^2+\beta^2)}{2}p_3^2=c_2^2$ & $$\\ \\
  
$\frac{g^{\prime2}}{(1-g^2)^2}$ & $\tanh{p_4r}$ & $\frac{1}{4}((\alpha-\beta)^2-4n(n+\alpha+\beta+1)$ & $c_3^2\tanh^2{p_4r}+c_4^2\tanh{p_4r}$ & $Nr^{-\frac{D-1}{2}}(\sech{p_4r})^{\frac{\alpha+\beta}{2}}$\\
$=p_4^2$ & $$ & $-2(\alpha+\beta))p_4^2$ & $-\frac{(D-1)(D-3)}{4r^2}$ & $\exp(-\frac{(\alpha-\beta)}{2}p_4r)P_n^{(\alpha,\beta)}(\tanh{p_4r})$\\ \\
 $$ & $$ & $$ & $with,\frac{1}{4}((2n+\alpha+\beta+1)^2-1)p_4^2=c_3^2;$ & $$\\
 $$ & $$ & $$ & $\hspace{1in}\frac{1}{2}(\alpha^2-\beta^2)p_4^2=c_4^2$ & $$ \\
\hline 
\end{tabular*}
\label{Tab3}
\end{sidewaystable}

\section{Conclusions}
\label{sec4}
In this article, we have presented a simple transformation method of construction of exactly solvable potentials using the properties of orthogonal polynomials in the regime of non-relativistic quantum mechanics. The method is applied to construct spherically symmetric exactly solvable potentials in arbitrary $D$-dimensional Euclidean space. The number of possible exactly solvable potentials that can be constructed using a particular orthogonal polynomial depends on the number of $g(r)$ dependent terms on the right hand side of equation (\ref{eq8}), the mode of extraction of energy eigenvalues as discussed in the formalism and the normalizability of the eigenfunctions. We have listed exactly solvable potentials constructed from associated Laguerre, Hypergeometric, associated Legendre and Jacobi polynomials. The method can however be applied to other orthogonal polynomials also. The constructed potentials are mostly non-power law with an inverse square potential $(D-1)(D-3)r^{-2}$ which vanishes for $D=1$ and $D=3$. For power law potential, this term along with Schwartzian derivative give the correct form of centrifugal barrier term in $D$-dimensions.(e.g.,equations (\ref{eq29}), (\ref{eq32})) It is notable that we have explicitly kept the various constants such as integration constants, scale factors and characteristic constants in our expressions, which allows flexibility to the constructed exactly solvable potentials at the time of possible applications.

\section*{Acknowledgements}
The authors thank (L)Prof. S A S Ahmed for his stimulating suggestions on the topic. One author (NB) thank UGC-RFSMS for financial support.
\\ \\ \\  \\ \\ \\ \\ \\ \\ \\ \\ \\ \\

\end{document}